           \documentclass[11pt,titlepage,a4paper]{article}
	   \usepackage{amsfonts}

\def\openone{\leavevmode\hbox{\small1\kern-3.8pt\normalsize1}}
\newtheorem{Th}{Theorem}[section]		
\newcommand{\QED}[1][6]{\hbox{\vrule width #1pt height #1pt depth 0pt}}
\newtheorem{Prop}{Proposition}[section]		
\newtheorem{Lem}{Lemma}[section]     		
\newtheorem{Cor}{Corollary}[section] 		
\newlength{\bo}\newlength{\ho}\newlength{\up}\newlength{\down}       
\newlength{\middle}                                                  
\newcommand{\bozho}{\leavevmode\hbox{\slshape\bfseries               
\settowidth{\bo}{BO}\settowidth{\ho}{HO}\settowidth{\middle}{/}
\settoheight{\up}{BOZHO}\settodepth{\down}{/}
\addtolength{\up}{+0.15\up}\addtolength{\bo}{+\middle}
\rule[\up]{\bo}{0.15ex}\hspace{-\bo}BO
\hspace{+0.09em}\raisebox{+0.17\up}{/}
\hspace{-0.20em}\raisebox{+0.71\up}{$\bullet$}
\hspace{-0.33em}\hspace{-1.14\middle}\raisebox{-0.4\up}{$\bullet$}
\hspace{-0.30em}\addtolength{\down}{-0.41\down}
\addtolength{\ho}{+1.5\middle}\rule[-\down]{\ho}{0.15ex}
\addtolength{\ho}{-\middle}\hspace{-\ho}\hspace{+0.18em}
\raisebox{+0.17\up}{HO}}}                                            
\newcommand{\BOZHO}
{\bozho$^{^{\text{\textregistered}\,} \text{\texttrademark} }$}      

\title{\bfseries	\vspace*{-1.23456789in}
\vspace*{-7ex}
{
\begin{flushright}
	  \textbf{\large LANL xxx E-print archive No. gr-qc/9805088}\\[5ex]
\end{flushright}
}
{\huge Normal frames and the validity of the equivalence principle}\\
\vspace{10pt}
{\LARGE III.  The case along smooth maps\\[4pt]
 with separable points of self-intersection} }
\author{
Bozhidar Z. Iliev\fnsymbol{footnote}
\thanks{Department Mathematical Modeling,
Institute for Nuclear Research and Nuclear Energy,
Bulgarian Academy of \mbox{Sciences},
blvd. Tzarigradsko Chauss\'ee 72, 1784 Sofia, Bulgaria}$^,$\thanks
{E-mail address: bozho@inrne.acad.bg}\\[6ex]
\textit{Short title: }
\textbf{\large Normal frames and the equivalence principle: III}
\\[3.6ex]
\textit{Classification numbers:} 02.40.Hw, 02.40.Ky, 04.20.Cv,
04.50.+h
}
\date{
 \vspace{5ex}
 Ended: December 26, 1992 \\
 Revised: July 4, October 19, 1996  \\
 Produced: \today \\[4pt]
 LANL xxx E-print archive No. gr-qc/9805088\\[5pt]
 Sent for submission to J. Phys. A :  May 12, 1997\\
 Sent for resubmission to J. Phys. A : November 12, 1997\\
 Published in
	J. Phys. A: Math. \& Gen. vol.31, No.~4, pp.1287--1296, 1998\\[7ex]
 {\Huge\bozho}
}

\begin{document}

\renewcommand{\thefootnote}{\fnsymbol{footnote}}
\maketitle
\renewcommand{\thefootnote}{\arabic{footnote}}

\tableofcontents

\begin{abstract}

The  equivalence principle is treated on a mathematically rigorous base on
sufficiently general subsets of a differentiable manifold. This is carried
out using the basis of derivations of the tensor algebra over that manifold.
Necessary  and/or  sufficient conditions of existence, uniqueness,  and
holonomicity of these bases in which the components of the derivations  of
the tensor algebra over it vanish on these subsets, are studied. The linear
connections  are considered in this context. It is shown that the equivalence
principle is identically valid at any point, and along any path, in every
gravitational theory based on linear connections. On higher dimensional
submanifolds it may be valid only in certain exceptional cases.

\end{abstract}

\section {Introduction}
\label{I}

In connection with the equivalence principle~\cite[ch. 16]{MTW}, as  well  as
from purely mathematical
reasons~\cite{Schouten/Ricci,K&N-I,Lovelock-Rund,Rashevskii},
an important problem is the existence of local
(holonomic or anholonomic~\cite{Schouten/Ricci})
coordinates (bases) in which the components of a linear connection~\cite{K&N-I}
vanish on  some subset, usually a submanifold, of a differentiable
manifold~\cite{K&N-I}.  This problem has been solved for torsion free, i.e.
symmetric,  linear connections~\cite{K&N-I,Lovelock-Rund} in the cases at a
point~\cite{Schouten/Ricci,K&N-I,Lovelock-Rund,Rashevskii}, along a smooth path
without self-intersections~\cite{Schouten/Ricci,Rashevskii},  and in a
neighborhood~\cite{Schouten/Ricci,Rashevskii}.  These results were
generalized in our previous
works~\cite{f-Bases-n+point,f-Frames-n+point,f-Bases-path,f-Frames-path} for
arbitrary, with or without torsion, derivations of  the tensor algebra over a
given differentiable manifold~\cite{K&N-I} and, in particular, for
arbitrary linear connections.  General results of this kind can be found
in~\cite{ORai}, where a criteria is presented for the existence of the
above-mentioned special bases (coordinates) on submanifolds of a space with a
symmetric affine connection.

The present work is a revised version of~\cite{f-Bases-general}
and is a continuation of~\cite{f-Frames-n+point,f-Frames-path}.
It generalizes the results
from~\cite{f-Frames-n+point,f-Frames-path,ORai} and
deals with the problems of  existence, uniqueness, and holonomicity of
special bases (frames) in which the components of a derivation of the tensor
algebra over a differentiable manifold vanish on some its subset of a
sufficiently general type (Sect.~\ref{II} and Sect.~\ref{III}). If such
frames exist, they are called \emph{normal}. In particular, the considered
derivation may be a linear connection (Sect.~\ref{IV}).  In this context we
make conclusions concerning the general validity and the mathematical
formulation of the equivalence principle  in  a class of gravitational
theories (Sect.~\ref{V}).

\section {\bfseries Mathematical preliminaries}
\label{IInew}
\setcounter {equation} {0}

Below we   reproduce for further reference purposes, as well  as  for   the
exact statement of the above problems, a  few simple facts about derivations
of tensor  algebras  that  can  be found
in~\cite{f-Frames-n+point,f-Frames-path} or derived from the those
in~\cite{K&N-I}.


Let $D$ be a derivation of the tensor algebra over a manifold
$M$~\cite{DNF-2,K&N-I}.
By~\cite[proposition 3.3 of chapter I]{K&N-I} there exist
a unique vector field $X$ and a unique tensor field $S$ of type $(1,1)$ such
that $D=L_{X}+S$. Here $L_{X}$ is the Lie derivative along
$X$~\cite{DNF-2,K&N-I}
and $S$ is considered as a derivation of the tensor algebra over
$M$~\cite{K&N-I}.

If $S$ maps from the set  of $C^{1}$  vector  fields  into  the
tensor fields of type (1,1) and $S:X\mapsto S_{X}$,  then  the equation
$ D^{S}_{X}=L_{X}+S_{X} $	
defines a derivation of the tensor algebra over $M$ for any $C^{1}$  vector
field $X$ \cite{K&N-I}.
Such a derivation will be called an $S$-derivation along $X$ and
denoted for brevity simply by $D_{X}$. An $S$-derivation is  a  map $D$  such
that $D:X\mapsto D_{X}$, where $D_{X}$ is an $S$-derivation along X.

Let $\{E_{i}, i=1,\ldots, n:=\dim (M)\}$ be a (coordinate or
not~\cite{Schouten/Ricci,Lovelock-Rund}) local basis (frame) of vector fields
in the tangent bundle to $M$. It is holonomic (anholonomic) if the vectors
$E_{1}, \ldots , E_{n}$ commute  (do  not
commute)~\cite{Schouten/Ricci,Lovelock-Rund}.
Using the explicit action of $L_{X}$  and $S_{X}$ on  tensor fields~\cite{K&N-I}
one can easily deduce the explicit form of the local components of $D_X T$
for any $C^1$ tensor field $T$. In particular,
the \emph{components} $(W_{X})^{i}_{j}$   of $D_{X}$ are defined by
	\begin{equation}	 \label{2}
D_{X}(E_{j})=(W_{X})^{i}_{j}E_{i}.
	\end{equation}
Here  and below all Latin indices, perhaps with some super- or subscripts,
run  from  $1$ to $n:=\dim (M)$ and
the usual summation rule on indices repeated  on  different levels is
assumed. It is easily seen that
$(W_{X})^{i}_{j}:=(S_{X})^{i}_{j}- E_{j}(X^{i}) + C^{i}_{kj}X^{k}$  
where
$X(f)$ denotes the action of $X=X^{k}E_{k}$ on  the $C^{1}$ scalar function
$f$, as $X(f):=X^{k}E_{k}(f)$,
and the $C^{i}_{kj}$ define the commutators of the basic
vector fields by
$[E_{j},E_{k}]=C^{i}_{jk}E_{i}$.	

The change $\{ E_{i}\} \mapsto \{ E^{\prime }_{m}:=A^{i}_{m}E_{i}\} $,
$A:=\left[ A^{i}_{m}\right]$  being  a
nondegenerate  matrix  function,  implies  the  transformation  of
$(W_{X})^{i}_{j}$ into (see (\ref{2}))
$(W^{\prime }_{X})^{m}_{l}=(A^{-1})^{m}_{i}A^{j}_{l}$
$(W_{X})^{i}_{j}+(A^{-1})^{m}_{i}X(A^{i}_{l})$. 	
Introducing the  matrices
$W_{X}:=[ (W_{X})^{i}_{j} ] $
and
$W^{\prime }_{X}:=[ (W^{\prime }_{X})^{m}_{l} ] $
and putting
$X(A):=X^{k}E_{k}(A)=[ X^{k}E_{k}(A^{i}_{m}) ] $, we get
	\begin{equation}	\label{5'}
W^{\prime }_{X}=A^{-1}\{W_{X}A+X(A)\}.
	\end{equation}

	If  $\nabla $ is a linear connection with local components
$\Gamma ^{i}_{jk}$
(see, e.g., \cite{DNF-2,K&N-I,Lovelock-Rund}), then
$\nabla _{X}(E_{j})=(\Gamma ^{i}_{jk}X^{k})E_{i}$~\cite{K&N-I}.	
Hence,  we see from (\ref{2}) that
$D_{X}$ is a covariant differentiation along $X$ iff
	\begin{equation}	\label{7}
(W_{X})^{i}_{j}=\Gamma ^{i}_{jk}X^{k}
	\end{equation}
for some functions $\Gamma ^{i}_{jk}$.

Let $D$ be an S-derivation and $X$ and $Y$ be vector fields.
The \emph{torsion operator} $T^{D}$ of $D$ is defined as
		\begin{equation}   \label{8}
T^{D}(X,Y):=D_{X}Y-D_{Y}X-[X,Y].
		\end{equation}
The S-derivation $D$ is \emph{torsion free} if $T^{D}=0$ (cf. \cite{K&N-I}).

For a linear connection $\nabla $, due to (\ref{7}), we have
%
$(T^{\nabla }(X,Y))^{i}     =  T^{i}_{kl}X^{k}Y^{l}$
%
where
 $ T^{i}_{kl}
    := - (\Gamma ^{i}_{kl} - \Gamma ^{i}_{lk}) - C^{i}_{kl}$
are the components  of the torsion tensor of $\nabla $~\cite{K&N-I}.


Further  we  investigate  the problem 	of existence of
bases $\{E_{i}^{\prime }\}$ in which $W_{X}^\prime =0$ for
an S-derivation $D$ along any or a fixed vector field X. These bases
(frames), if any, are called \emph{normal}. Hence, due to~(\ref{5'}), we
have to solve the equation $W_{X}(A)+X(A)=0$ with respect to $A$ under
conditions that will be presented below.

\section {\bfseries Derivations along every vector fields}
\label{II}
\setcounter {equation} {0}

This section is devoted to the existence and
some properties of special bases
(frames) $\{E_{i}^{\prime }\}$, defined in a neighborhood of a subset $U$ of
the manifold $M$, in which the components of an $S$-derivation $D_X$ along an
\emph{every} vector field $X$ vanish on $U$.  These bases (frames), if
any, are called \emph{normal in}  $U$.

The derivation $D$ is called \emph{linear on  the set} $U\subseteq M$ if
(cf.~(\ref{7})) in  some (and hence  in  any) basis $\{E_{i}\}$ is fulfilled
	\begin{equation}	\label{10}
W_{X}(x)=\Gamma _{k}(x)X^{k}(x),
	\end{equation}
where $x\in U$, $X=X^{k}E_{k}$, and $\Gamma _{k}$ are some matrix functions
on $U$. Evidently, a  linear connection on $M$ is  a linear on $U$ for every
$U$ (see~(\ref{7})).

	\begin{Prop}	\label{Prop1}
If for some $S$-derivation $D$  there  exists  a normal basis
$\{E_{i}^{\prime }\}$ in $U\subseteq M$, i.e.
$\left.W_{X}^\prime\right|_{U}=0$ for every vector field $X$,
then $D$  is  linear on the set $U$.
	\end{Prop}

\textit{Proof.}  Let us fix  a  basis $\{E_{i}\}$ and put
$E_{i}^{\prime }=A^{j}_{i}E_{j}$. Then
$\left. W_{X}^\prime\right|_{U}=0$, i.e.
$W_{X}^\prime(x)=0$ for $x\in U$,
which, in conformity  with~(\ref{5'}), is  equivalent to~(\ref{10}) with
$\Gamma _{k}=-(E_{k}(A))A^{-1}$,
$A=[ A^{i}_{j} ]$.~\QED

The opposite statement to proposition~\ref{Prop1} is generally not true  and
for its appropriate formulation  we need   some   preliminary results and
explanations.

Let $p$  be an integer, $p\ge 1,$ and the Greek indices $\alpha $ and
$\beta $ run from 1 to $p$. Let $J^{p}$ be a neighborhood in $\mathbb{R}^{p}$
and $\{s^{\alpha }\}=\{s^{1},\ldots ,s^{p}\}$ be (Cartesian) coordinates in
$\mathbb{R}^{p}$.

	\begin{Lem}	\label{Lem2}
Let
$Z_{\alpha }:J^{p}\to \mathrm{GL}(m,\mathbb{ R})$,
$\mathrm{GL}(m,\mathbb{R})$ being the group of $m\times m$
matrices  on $\mathbb{R}$,  be $C^{1}$  matrix-valued functions on $J^{p}$.
Then the initial-value problem
	\begin{equation}	\label{11}
\left.\frac{\partial Y}{\partial s^{\alpha }}\right|_{s}=Z_{\alpha }(s)Y,
\quad
\left.Y\right|_{s=s_{0}}=\openone,
\quad \alpha=1,\ldots,p,
      	\end{equation}
where
${\openone:= \left[ \delta ^{i}_{j}\right]^{m}_{i,j=1}}$
is  the  unit matrix of the corresponding size,
$s\in J^{p}$, $s_{0}\in J^{p}$  is fixed, and $Y$ is $m\times m$ matrix
function on $J^{p}$, has a solution,
denoted by $Y=Y(s,s_{0};Z_{1},\ldots ,Z_{p})$, which is unique  and  smoothly
depends on all its arguments if and only if
	\begin{equation}	\label{12}
R_{\alpha \beta }(Z_{1},\ldots ,Z_{p}):=
\frac{ \partial Z_{\alpha } }{ \partial s^{\beta } } -
\frac{ \partial Z_{\beta }  }{\partial s^{\alpha }  } +
Z_{\alpha }Z_{\beta } - Z_{\beta } Z_{\alpha } =0.
	\end{equation}
	\end{Lem}
\textit{Proof.}  According  to the results from~\cite[chapter~VI]{Hartman},
in which $Z_{1},\ldots ,Z_{p}$ are of class $C^1$, the
integrability conditions for~(\ref{11}) are
(cf.~\cite[chapter~VI, equation~(1.4)]{Hartman})
	\begin{eqnarray*}
0 & = &
\frac{\partial ^{2}Y} {\partial s^{\alpha }\partial s^{\beta }  } -
    \frac{\partial ^{2}Y} {\partial s^{\beta }\partial s^{\alpha }  } =
\frac{\partial (Z_{\beta }Y) } { \partial s^{\alpha } } -
\frac{\partial (Z_{\alpha }Y)} {\partial s^{\beta }   } \> =  \\
& = &
\frac{ \partial Z_{\beta }  }{\partial s^{\alpha } } Y -
\frac{ \partial Z_{\alpha } }{\partial s^{\beta } }Y +
Z_{\beta }Z_{\alpha }Y - Z_{\alpha }Z_{\beta }Y =
- R_{\alpha \beta }(Z_{1},\ldots ,Z_{p}) Y.
	\end{eqnarray*}
Hence (see, e.g.~\cite[chapter~VI, theorem~6.1]{Hartman}) the
initial-value problem~(\ref{11}) has a unique solution (of class $C^2$)
iff~(\ref{12}) is satisfied.~\QED

Let $p\le n:=\dim (M)$, $\alpha ,\beta =1,\ldots,p$ and
$\mu ,\nu =p+1,\ldots,n$. Let $\gamma :J^p\to M$
be a $C^{1}$  map. We suppose that for any $s\in J^{p}$ there exists
its ($p$-dimensional)  neighborhood $J_{s}\subseteq J^{p}$  such that
the restricted map $\left. \gamma\right|_{J_{s}}:J_{s}\to M$ is without
self-intersections, i.e. in $J_{s}$ does not
exist  points $s_{1}$  and $s_{2}\neq s_{1}$  with  the property
$\gamma (s_{1})=\gamma (s_{2})$. This
assumption   is   equivalent   to   the  one  that  the  points  of
self-intersections  of $\gamma $, if any, can be separated by neighborhoods.
With $J^{p}_{s}$ we denote the union of all the neighborhoods $J_{s}$ with
the above  property;  evidently, $J^{p}_{s}$ is the maximal neighborhood of
$s$ in which $\gamma $ is without self-intersections.

	Let us suppose  at  first that  $J^{p}_{s}=J^{p}$,  i.e. that
$\gamma $ is without self-intersection,  and  that $\gamma (J^{p})$  is
contained in a single coordinate neighborhood $V$ of $M$.

Let  us  fix  some one-to-one $C^{1}$ map $\eta :J^{p}\times J^{n-p}\to M$
such that
$\eta (\cdot,\mathbf{ t}_{0}) = \gamma$ for a fixed
$\mathbf{ t}_{0}\in J^{n-p}$,
i.e. $\eta (s,\mathbf{ t}_{0})=\gamma (s)$, $s\in J^{p}$.
In $V\bigcap \eta (J^{p},J^{n-p})$ we  define  coordinates
$\{x^{i}\}$ by putting
\(
(x^{1}(\eta (s,\mathbf{ t})), \ldots,x^{n}(\eta (s,\mathbf{ t}))):=
(s,\mathbf{ t})\in \mathbb{R}^{n}
\),
$s\in J^{p}$, $\mathbf{ t}\in J^{n-p}$.

	\begin{Prop}	\label{Prop3}
Let $\gamma :J^{p}\to M$ be a $C^{1}$ map without self-intersections and such
that $\gamma (J^{p})$ lies only in one coordinate neighborhood. Let the
derivation $D$  be linear on $\gamma (J^{p})$. Then a necessary and
sufficient condition for the existence of a basis $\{E_{i}^{\prime }\}$,
defined  in  a  neighborhood of $\gamma (J^{p})$,
in which the components  of $D$
along every vector field vanish on $\gamma (J^{p})$ is the validity in the
above-defined coordinates $\{x^{i}\}$ of the equalities
	\begin{equation}	\label{13}
\left.
[R_{\alpha \beta }(-\Gamma _{1}\circ \gamma ,\ldots,-\Gamma _{p}\circ\gamma)]
\right|_{J^{p}}=0, \qquad \alpha ,\beta =1,\ldots,p,
	\end{equation}
where $R_{\alpha \beta }(\ldots)$  are defined  by~(\ref{12}) for $m=n$
and $(s^{1},\ldots,s^{p})=s\in J^{p}$, i.e.
\begin{equation}	\label{14}
[R_{\alpha \beta }(-\Gamma _{1}\circ \gamma ,\ldots,
-\Gamma _{p}\circ\gamma)](s) \!=\!
\frac{\partial \Gamma _{\alpha }(\gamma (s))} {\partial s^{\beta } } -
\frac{\partial \Gamma _{\beta }(\gamma (s)) } {\partial s^{\alpha }} +
\left.
(\Gamma _{\alpha }\Gamma _{\beta } - \Gamma _{\beta }\Gamma _{\alpha })
\right|_{\gamma (s)}.
	\end{equation}
        \end{Prop}

\textbf{Remark.} This result was obtained by means of another method
in~\cite{ORai} for the special case when $D$ is a  symmetric  affine
connection and $U$ is a submanifold of  $M$.

\textit{Proof.} The following considerations will be done in the
above-defined  neighborhood $V\bigcap \eta (J^{p},J^{n-p})$  and
coordinates $\{x^{i}\}$.  Let $E_{i}=\partial/\partial x^{i}$.

\textsl{NECESSITY.}
Let there exists a  normal frame
$\{E_{i}^{\prime }=A^{j}_{i}E_{i}\}$ on $\gamma(J^p)$, i.e.
$W_{X}^\prime(\gamma (s))=0$, $s\in J^{p}$. By~(\ref{5'})  the existence
of $\{E_{i}^{\prime }\}$ is  equivalent to  that of
$A=[ A^{j}_{i} ]$, transforming $\{E_{i}\}$ into
$\{E_{i}^{\prime }\}$, and such that
$\left. [A^{-1}(W_{X}A+X(A))]\right|_{\gamma (s)}=0$ for \textit{every} X.
As $D$  is linear on  $\gamma (J^{p})$ (cf. proposition~\ref{Prop1}),
the equation~(\ref{10}) is valid for $x\in \gamma (J^{p})$ and some
matrix-valued functions $\Gamma _{k}$ .
Consequently $A$  must  be  a  solution  of
$\Gamma^\prime_{k }(x)=0$, i.e. of
	\begin{equation} 	\label{15}
\left.
\Gamma _{k}(\gamma (s))A(\gamma (s)) + \frac{\partial A}{\partial x^{k}}
\right|_{\gamma (s)}=0, \qquad s\in J^{p}.
	\end{equation}


	Now define nondegenerate matrix-valued functions $B$ and $B_i$ by
\[
A(\gamma (s)) = B(s), \qquad
\left. \frac{\partial A}{\partial x^{\alpha }} \right|_{\gamma (s)} =
\frac{\partial B(s)}{\partial s^{\alpha }}, \quad \alpha =1,\ldots,p,
\]
\[
\left.\frac{\partial A}{\partial x^{\nu }}\right|_{\gamma (s)} =
B_{\nu }(s), \qquad \nu =p+1,\ldots ,n.
\]
Substituting these equalities into~(\ref{15}), we see that it splits to
	\begin{eqnarray}	\label{17a}
\Gamma _{\alpha }(\gamma (s))B(s) & + &
\frac{\partial B(s)}{\partial s^{\alpha } }=0, \qquad \alpha =1,\ldots ,p,
\\	\label{17b}
\Gamma _{\nu }(\gamma (s))B(s) & + & B_{\nu }(s)=0,
\qquad \nu =p+1,\ldots ,n.
	\end{eqnarray}

As these equations do not involve $B_{\alpha }$, the $B_{\alpha }$'s are
left arbitrary by~(\ref{15}), while the remaining $B_{i}$'s are expressed
via $B(s)$ through (see ~(\ref{17b}))
	\begin{equation}	\label{18}
B_{\nu }(s)=-\Gamma _{\nu }(\gamma (s))B(s), \qquad \nu =p+1,\ldots,n.
	\end{equation}

	So,  $B(s)$ is the only quantity for determination. It must
satisfy~(\ref{17a}). If  we arbitrary fix the value $B(s_{0})=B_{0}$ for a
fixed $s_{0}\in J^{p}$  and put $Y(s)=B(s)B^{-1}_{0}$
($B$ is a nondegenerate as $A$ is such by definition),
we  see  that $Y$ is a solution of the initial-value problem
	\begin{equation}	\label{19}
\left. \frac{\partial Y}{\partial s^{\alpha } }\right|_{s} =
- \Gamma _{\alpha }(\gamma (s))Y(s), \qquad
\alpha =1,\ldots p, \quad
\left.Y\right|_{s=s_{0}} = \openone_{p} =
\left[\delta ^{i}_{j}\right]^{p}_{i,j=1}.
	\end{equation}

By lemma~\ref{Lem2}  this  initial-value problem has a unique solution
$ Y=Y(s,s_{0};-\Gamma _{1}\circ \gamma ,\ldots,-\Gamma _{p}\circ\gamma)$
iff the integrability conditions~(\ref{13}) are valid.

	Consequently the existence of  $\{ E_{i}^{\prime} \}$ (or of $A$)
leads to ~(\ref{13}).

	\textsl{SUFFICIENCY.}
If~(\ref{13}) take place, the general solution of~(\ref{17a}) is
	\begin{equation}	\label{20}
B(s) =
Y(s,s_{0};-\Gamma _{1}\circ \gamma ,\ldots,-\Gamma _{p}\circ \gamma )B_{0},
	\end{equation}
in  which $s_{0}\in J^{p}$ and the nondegenerate matrix $B_{0}$ are fixed.
Consequently, admitting $A$ to be a $C^1$ matrix-valued function, we see that
in $V\bigcap \eta (J^{p},J^{n-p})$ we can expand
$A(\eta (s,\mathbf{ t}))$, $s\in J^{p}$, $\mathbf{ t}\in J^{n-p}$
up to second order terms with respect to $(\mathbf{ t}-\mathbf{ t}_{0})$ as
	\begin{eqnarray} 	\nonumber
&  A  (\eta (s,\mathbf{ t}))  =
B(s)+B_{i}(s)[x^{i}(\eta (s,\mathbf{ t}))-x^{i}(\eta (s,\mathbf{ t}_{0}))]
\> + &
\\ 	\label{16}
  & + \>
B_{ij}(s,\mathbf{ t};\eta )
[x^{i}(\eta (s,\mathbf{ t}))-x^{i}(\eta (s,\mathbf{ t}_{0}))]
[x^{j}(\eta (s,\mathbf{ t}))-x^{j}(\eta (s,\mathbf{ t}_{0}))] &
	\end{eqnarray}
for the above-defined matrix-valued functions $B,\ B_{i}$, and some $B_{ij}$,
which are such that  $\det B(s)\not = 0,\infty$ and  $B_{ij}$ and their first
derivatives are bounded when $\mathbf{ t}\to \mathbf{ t}_{0}$.  (Note that
in~(\ref{16}) the terms corresponding to $i,j=1,\ldots ,p$ are equal to zero
due to the definition of $\{x^{i}\}$.)
In this case, due to~(\ref{17a})--(\ref{20}), the general solution
of~(\ref{15}) is
	\begin{eqnarray}	\nonumber
&
A(\eta (s,\mathbf{ t})) =
\left\{
\openone -
\sum^{n}_{\lambda =p+1}\Gamma _{\lambda }(\gamma (s))
[x^{\lambda }(\eta (s,\mathbf{ t}))-x^{\lambda }(\gamma (s))]
\right\} \> \times
&
\\ \nonumber
&
\times\> Y(s,s_{0};-\Gamma _{1}\circ \gamma,\ldots,-\Gamma_{p}\circ \gamma)
B_{0} \> +
\sum^{n}_{\mu ,\nu =p+1}
\Bigl\{
B_{\mu \nu }(s,\mathbf{ t};\eta ) \times
&
\\	\label{21}
&
\times [x^{\mu }(\eta (s,\mathbf{ t}))  -
x^{\mu }(\gamma (s))][x^{\nu}(\eta(s,\mathbf{ t}))-x^{\nu}(\gamma(s))]
\Bigr\}  ,
&   \!\!\!\!\!  \!\!\!\!\!
	\end{eqnarray}
where $s_{0}\in J^{p}$ and  the  nondegenerate  matrix $B_{0}$ are fixed and
$B_{\mu ,\nu }$, $\mu ,\nu =p+1,\ldots,n$,  together  with  their first
derivatives are bounded when $\mathbf{ t}\to \mathbf{ t}_{0}$.
(The fact that into~(\ref{21}) enter only sums from $p+1$ to $n$ is a
consequence from
$x^{\alpha }(\eta (s,\mathbf{ t}))=x^{\alpha }(\gamma (s))=s^{\alpha }$, i.e.
$x^{\alpha }(\eta (s,\mathbf{ t})) - x^{\alpha }(\eta (s,\mathbf{ t}_{0})) =
x^{\alpha }(\eta (s,\mathbf{ t}))-x^{\alpha }(\gamma (s)) =
s^{\alpha } - s^{\alpha }\equiv 0$,
$\alpha = 1,\ldots,p$.)

Hence, from~(\ref{13}) follows the existence of a class of matrices $A(x)$,
$x\in V \bigcap  \eta(J^p,J^{n-p})$ such that the frames
$\{ E_{i}^{\prime}=A_{i}^{j}E_j \}$ are normal for $D$ (which is supposed to
be linear on $\gamma(J^p)$).~\QED

Thus  bases $\{E_{i}^{\prime }\}$ in which $W_{X}^\prime=0$ exist
iff~(\ref{13}) is satisfied.  If~(\ref{13}) is valid, then the normal bases
$\{E_{i}^{\prime }\}$ are obtained from $\{E_{i}=\partial/ \partial x^{i}\}$
by means of linear transformations whose matrices must  have the
form~(\ref{21}).

Now we are  ready  to  consider  a  general  smooth $(C^{1})$  map
$\gamma :J^{p}\to M$ whose points of self-intersection, if any,  can  be
separated by neighborhoods. For any $r\in J^{p}$  chose  a
 coordinate neighborhood $V_{\gamma (r)}$ of $\gamma (r)$ in $M$. Let there
be given  a  fixed $C^{1}$ one-to-one map
$\eta_{r}:J^{p}_{r}\times J^{n-p}\to M$ such that
$\eta _{r}(\cdot ,\mathbf{ t}^{r}_{0})=\left.\gamma\right|_{J^{p}_{r}}$  for
some $\mathbf{ t}^{r}_{0}\in J^{n-p}$. In the neighborhood
$V_{\gamma (r)}\bigcap \eta _{r}(J^{p}_{r},J^{n-p})$  of
$\gamma (J^{p}_{r})\bigcap V_{\gamma (r)}$  we
introduce local coordinates $\{x^{i}_{r}\}$ defined by
\[
(x^{1}_{r}(\eta _{r}(s,\mathbf{ t})),\ldots,x^{n}_{r}
(\eta _{r}(s,\mathbf{ t}))) := (s,\mathbf{ t})\in \mathbb{R}^{n},
\]
where
$ s\in J^{p}_{r}\hbox{ and }\mathbf{ t}\in J^{n-p}$ are such that
$ \eta _{r}(s,\mathbf{ t})\in V_{\gamma (r)}$.

	\begin{Th}	\label{Th4}
Let the points of self-intersection of  the $C^{1}$  map
$\gamma :J^{p}\to M$, if any, be separable by neighborhoods. Let
the $S$-derivation $D$ be linear on $\gamma (J^{p})$ ,  i.e.~(\ref{10})
to  be valid for  $x\in\gamma(J^p)$. Then a necessary and sufficient
condition for the existence in some neighborhood of $\gamma (J^{p})$ of a
basis $\{E_{i}^{\prime }\}$ in which the components of $D$ (along every
vector field) vanish on $\gamma (J^{p})$ is for every $r\in J$ in the
above-defined local coordinates $\{x^{i}_{r}\}$ to be fulfilled
	\begin{equation}	\label{22}
\left[ R_{\alpha \beta }
(-\Gamma _{1}\circ \gamma ,\ldots,-\Gamma _{p}\circ \gamma )\right](s)=0,
\qquad  \alpha ,\beta =1,\ldots,p,
	\end{equation}
where $\Gamma _{\alpha }$ are calculated by means of~(\ref{10}) in
$\{x^{i}_{r}\}, R_{\alpha \beta }$ are given by~(\ref{14}), and
$s\in J^{p}_{r}$ is such that $\gamma (s)\in V_{\gamma (r)}$.
	\end{Th}

\textit{Proof.}  For any $r\in J^{p}$ the restricted map
$\left.\gamma\right|_{{^\prime\!} J^{p}_{r}}:{^\prime\!} J^{p}_{r}\to M$,
where
${^\prime\!}J^{p}_{r}:=\{s\in J^{p}_{r},\  \gamma (s)\in V_{\gamma (s)}\}$,
is  without  self-intersections (see the above  definition  of
$J^{p}_{r})$ and
\(
\left.\gamma\right|_{
{^\prime\!} J^{p}_{r}}( {^\prime\!} J^{p}_{r}) =
\gamma ({^\prime\!} J^{p}_{r})
\)
lies in the coordinate neighborhood $V_{\gamma (r)}$.

So, if exists a normal frame $\{E_{i}^{\prime }\}$ for $D$, then, by
proposition~\ref{Prop3}, the equations~(\ref{22}) are identically satisfied.

Conversely,  if~(\ref{22})  are  valid,  then,  again,  by
proposition~\ref{Prop3} for  every $r\in J^{p}$ in a certain neighborhood
${^\prime} V_{r}$ of
$\gamma ({^\prime\!} J^{p}_{r})$ in $V_{\gamma (r)}$  exists
a  normal on $\gamma ( {^\prime\!} J^{p}_{r})$  basis
$\{E^{r}_{i}\}$  for $D_{X}$ along every vector  field $X$.
From the neighborhoods ${^\prime} V_{r}$ we  can  construct
a  neighborhood $V$  of $\gamma (J^{p})$, e.g., by putting
$V=\bigcup_{r\in J^p} {^\prime}V_{r}$.
Generally, $V$ is  sufficient to be taken as a  union of ${^\prime}V_{r}$ for
some, but not all $r\in J^{p}$. On $V$ we can obtain a normal basis
$\{E_{i}^{\prime }\}$ by putting
$\left.E_{i}^{\prime }\right|_{x} = \left.E^{r}_{i}\right|_{x}$ if
$x$ belongs to only one neighborhood ${^\prime} V_{r}$. If $x$ belongs to
more than one neighborhood ${^\prime} V_{r}$ we can choose
$\{\left.E_{i}^{\prime }\right|_{x}\}$ to be the basis
$\{\left.E^{r}_{i}\right|_x\}$ for some arbitrary fixed $r$.~\QED

	\textbf{Remark.} Note that generally the basis obtained at the end of
the proof of theorem~\ref{Th4} is not continuous in the regions containing
intersections of several neighborhoods ${^\prime} V_{r}$. Hence it is,
generally, no longer differentiable there. Therefore the adjective `normal' is
not very suitable in the mentioned regions. May be in such cases is better to
be spoken about `special' frames instead of `normal' ones.

	\begin{Prop}	\label{Prop5}
If on the set $U\subseteq M$ there exists normal frames on $U$ for some
$S$-derivation along every vector field, then all of them are connected
by  linear transformations whose coefficients are such that the
action on them of the corresponding basic vectors vanishes on U.
	\end{Prop}

\textit{Proof.}  If $\{E_{i}\}$  and
$\{E_{i}^{\prime }=A^{j}_{i}E_{j}\}$ are normal on $U$ bases, i.e. if
$W_{X}(x)=W_{X}^\prime(x)=0$  for $x\in U$
and every vector field $X=X^{i}E_{i}$, then due to~(\ref{5'}), we have
$\left.X(A)\right|_{U}=0,$ i.e. $\left.E_{i}(A)\right|_{U}=0.$ Conversely, if
$\left.W_{X}\right|_{U}=0$ in $\{E_{i}\}$   and
$E_{i}^{\prime } = A^{j}_{i}E_{j}$   with
$\left.E_{i}(A)\right|_{U}=0,$  then
from~(\ref{5'}) follows
$\left.W_{X}^\prime(x)\right|_{U}=0,$  i.e. $\{E_{i}^{\prime }\}$ is
also a normal basis.~\QED

	\begin{Prop}	\label{Prop6}
If for some $S$-derivation $D$ there exists a
local holonomic normal basis on the
set $U\subseteq M$ for $D$ along every  vector field, then $D$ is torsion
free on  $U$.  On  the other hand, if $D$ is torsion free on $U$ and there
exist smooth ($C^1$) normal bases on  $U$ for $D$ along every vector field,
then all of them are holonomic on $U$, i.e. their basic vectors
commute on $U$.

	\end{Prop}

\textit{Proof.}  If $\{E_{i}^{\prime }\}$ is a normal basis on  $U$, i.e.
$W_{X}^\prime(x)=0$  for  every $X$ and $x\in U$, then using~(\ref{2})
and~(\ref{8}) (see also~\cite[eq. (15)]{f-Bases-n+point}), we find
\(
\left. T^{D}(E_{i}^{\prime },E_{j}^{\prime })\right|_{U} =
- \left. [E_{i}^{\prime},E_{j}^{\prime }]\right|_{U}.
\)
Consequently
$\{E_{i}^{\prime }\}$ is holonomic on $U$, i.e.
$\left. [E_{i}^{\prime },E_{j}^{\prime }]\right|_{U}=0$, iff
$0=\left.T^{D}(X,Y)\right|_{U} = \left.
\{X^{{\prime }i}Y^{{\prime }j}
T^{D}(E_{i}^{\prime },E_{j}^{\prime })\}\right|_{U}$
for  every  vector fields $X$ and $Y$, which is equivalent to
$\left.T^{D}\right|_{U}=0$.

Conversely, let $\left.T^{D}\right|_{U}=0.$ We want to prove that any
basis $\{E_{i}^{\prime }=A_{i}^{j}E_j\}$  in which $W_{X}^\prime=0$ is
holonomic on U.  The holonomicity on $U$ means
\(
0=\left.[E_{i}^{\prime },E_{j}^{\prime }]\right|_{U} =
\left. \{- (A^{-1})^{l}_{k}
[ E_{j}^{\prime }(A^{k}_{i }) - E_{i}^{\prime }(A^{k}_{j }) ]
E_{l}^{\prime }\} \right|_{U}.
\)
However (see proposition~\ref{Prop1} and~(\ref{10})) the existence of
$\{E_{i}^{\prime }\}$ is equivalent to
$\left.W_{X}\right|_{U}=\left.(\Gamma _{k}X^{k})\right|_{U}$
for  some  functions $\Gamma _{k}$ and every X. These two facts,
combined with~(\ref{2}) and~(\ref{8}),  lead to
$(\Gamma _{k})^{i}_{ j}=(\Gamma _{j})^{i}_{ k}$.  Using  this  and
$\left. \{\Gamma _{k}A + \partial A/\partial x^{k}\}\right|_{U}=0$
(see the proof of proposition~\ref{Prop1}), we find
\(
\left. E_{j}^{\prime }(A^{k}_{i })\right|_{U} =
- \left.\left\{ A^{l}_{j }A^{m}_{i }
(\Gamma _{l})^{k}_{ m} \right\}\right|_{U} =
\left.\left( E_{i}^{\prime }(A^{k}_{j }) \right)\right|_{U}.
\)
Therefore
$\left. [E_{i}^{\prime },E_{j}^{\prime }]\right|_{U}=0$ (see above), i.e.
$\{E_{i}^{\prime }\}$ is holonomic on~U.~\QED

\section {\bfseries Derivations along a fixed vector field}
\label{III}
\setcounter {equation} {0}

	In this section we briefly outline some results concerning normal
frames for (S-)derivations along a \emph{fixed vector field}.

	A derivation $D_X$ is linear on $U\subseteq M$ along a \emph{fixed}
vector field $X$  if~(\ref{10}) holds for $x\in U$  and the given $X$. In this
sense, evidently,
\emph{any derivation along a fixed vector field is linear on every set}
and, consequently, on the whole manifold $M$. Namely this is the cause due to
which the analogue of proposition~\ref{Prop1} for such derivations, which is
evidently true, is absolutely trivial and does even need not to be formulated.


The existence of normal frames in which the components of $D_{X}$,  with  a
\emph{fixed} $X$, vanish on some set $U\subseteq M$ significantly differs
from the  same  problem  for $D_{X}$  with an \textit{every} $X$
(see Sect.~\ref{II}). In fact,
if $\{E_{i}^{\prime }=A^{j}_{i }E_{j}\}$,
$\{E_{i}\}$ being  a fixed basis on $U$, is a normal frame on $U$, i.e.
$\left.W_{X}^\prime\right|_{U}=0,$ then, due to~(\ref{5'}), its
existence is equivalent  to  the one  of $A:=[ A^{j}_{i } ]$
for which $\left.(W_{X}A+X(A))\right|_{U}=0$ for the \textit{given}  X.  As
$X$ is \textit{fixed}, the values of A at two different points,
say $x,y\in U$, are connected through the last equation if and only if $x$
and $y$ lie on one and the same integral curve of $X$, the part of
which between $x$ and $y$ belongs entirely to $U$. Hence, if $\gamma:J\to M$,
$J$ being an $\mathbb{R}$-interval, is (a part of) an integral curve of $X$,
i.e.  at $\gamma (s), s\in J$ the tangent to $\gamma $ vector field
$\dot{\gamma }$ is $\dot{\gamma }(s):=\left.X\right|_{\gamma (s)}$, then
along $\gamma$ the equation $\left.(W_{X}A+X(A))\right|_{U}=0$ reduces to
\(
\left.  dA/ds\right|_{\gamma (s)}=\left. \dot{\gamma }(A)\right|_{s} =
\left.(X(A))\right|_{\gamma (s)} = - W_{X}(\gamma (s))A(\gamma (s)).
\)
Using lemma~\ref{Lem2} for $p=1$, we see that the  general solution of this
equation is
	\begin{equation}	\label{23}
A(s;\gamma ) = Y(s,s_{0};-W_{X}\circ \gamma )B(\gamma ),
	\end{equation}
where $s_{0}\in J$ is fixed, $Y=Y(s,s_{0};Z)$, $Z$ being a $C^{1}$
matrix function of $s$, is the unique solution of the initial-value problem
(see~\cite[ch. IV, \S 1]{Hartman})
	\begin{equation}	\label{24}
\frac{dY}{ds}=ZY, \quad \left. Y \right|_{s=s_{0}} = \openone,
	\end{equation}
and  the nondegenerate matrix $B(\gamma )$ may depend only on $\gamma $, but
not on s. (Note that~(\ref{24}) is a special case of~(\ref{11}) for
$p=1$ and by lemma~\ref{Lem2} it has always a unique solution because
$R_{11}(Z_{1})\equiv 0$ due to ~(\ref{12}) for $p=1$.)

From the above considerations, the next propositions follow.

	\begin{Prop}	\label{Prop9}
      	There exist normal bases for any S-derivation along a fixed vector
field on every set $U\subseteq M$.
	\end{Prop}

	\begin{Prop}	\label{Prop8}
The normal on the set $U\subseteq M$ bases  for some S-derivation along a
fixed vector field  $X$ are connected by linear transformations
whose matrices are such that the action  of $X$ on them vanishes on~$U$.
	\end{Prop}

\textit{Proof.}
If $\{E_{i}\}$ and $\{E_{i}^{\prime }=A^{j}_{i}E_{j}\}$ are such that
$\left.W_{X}^\prime\right|_{U} = \left.W_{X}\right|_{U}=0$, then, due
to~(\ref{5'}), we have $\left.X(A)\right|_{U}=0.$ On the other hand,
if $\left.W_{X}\right|_{U}=0$ and $\left.X(A)\right|_{U}=0,$  then,
by~(\ref{5'}), is fulfilled $\left.W_{X}^\prime\right|_{U}=0,$
i.e.  $\{E_{i}^{\prime }\}$ is normal.~\QED

\section {\bfseries Linear connections}
\label{IV}
\setcounter {equation} {0}

The  results  of Sect.~\ref{II} can directly be applied to the case of
linear connections.  As this is more or less trivial, we present below only
three such consequences.

	\begin{Cor}	\label{Cor10}
Let  the points of self-intersection of the $C^{1}$  map
$\gamma :J^{p}\to M$,
if any, be separable by neighborhoods, $\nabla $ be a linear connection  on
$M$ with local components $\Gamma ^{i}_{ jk}$ (in a basis $\{E_{i}\})$ and
$\Gamma _{k} := \left[ \Gamma ^{i}_{ jk} \right]^{n}_{i,j=1}$.
Then in a neighborhood of $\gamma (J^{p})$ there exists a normal
frame $\{E_{i}^{\prime }\}$ on
$\gamma (J^{p})$  for $\nabla $, i.e.
$\left. \Gamma _{k}^{\prime } \right|_{\gamma (J^{p})}=0$,  iff  for  every
$r\in J^{p}$  in the coordinates $\{x^{i}_{r}\}$ (defined before
theorem~\ref{Th4}) is satisfied~(\ref{22}) in which $\Gamma _{\alpha }$,
$\alpha =1,\ldots ,p$ are part of the components of $\nabla $ in
$\{x^{i}_{r}\}$ and $s\in J^{p}$ is such that $\gamma (s)\in V_{\gamma (r)}$.
	\end{Cor}

\textit{Proof.}  For linear connections~(\ref{10}) is valid for every $X$
in  any  basis. So, if in a basis $\{E_{i}^{\prime }\}$ is fulfilled
$\left. W_{X}^\prime \right|_{U}=0$ for $U\subseteq M$,  we  have in it
$\left. \Gamma _{k}^{\prime } \right|_{U}=0$ (see~(\ref{5'})) and vice versa,
if in a basis $\{E_{i}^{\prime }\}$ is valid
$\left. \Gamma _{k}^{\prime }\right|_{U}=0$,  then
$\left. W_{X} \right|_{U}=0$ for every X. Combining this fact with
theorem~\ref{Th4}, we get the required result.~\QED

	\begin{Cor}	\label{Corr11}
	If on the set $U\subseteq M$ there exist normal frames for some linear
connection on $U$, then these frames are connected
by linear transformations whose  matrices are such that
the action  of  the  corresponding basic vectors on them vanishes on $U$.
	\end{Cor}

\textit{Proof.}  The  result follows from proposition~\ref{Prop5} and the
proof of corollary~\ref{Cor10}.~\QED

	\begin{Cor}	\label{Corr12}
Let, for some linear connection on a neighborhood  of some set
$U\subseteq M$, there exist locally smooth normal bases on $U$. Then one
(and hence any) such basis is holonomic on $U$ iff the connection is
torsion free on $U$.
	\end{Cor}

\textit{Proof.}  The statement follows from~(\ref{10}) (or~(\ref{7})) and
proposition~\ref{Prop6}.~\QED

\section {\bfseries Conclusion. The equivalence principle}
\label{V}
\setcounter {equation} {0}

Mathematically theorem~\ref{Th4}  is the main result of this work.
From the view point of its physical application, it expresses  a
sufficiently general necessary and sufficient condition for existence
of the considered here normal frames for tensor derivations,  that, in
particular, can be linear connections. For instance, it covers that
problem  on  arbitrary submanifolds. In this sense, its special cases
are the results
from~\cite{ORai} and from our previous
papers~\cite{f-Frames-n+point,f-Frames-path}.

	Let $\gamma\colon J^p\to M$, with $J^p$ being a neighborhood in
$\mathbb{R}^p$ for some integer $p\le\dim{M}$, be a  $C^1$ map. If $p=0$  or
$p=1,$ then the conditions~(\ref{22}) are identically satisfied, i.e.
$R_{\alpha \beta }=0$ (see~(\ref{14})). Hence in these two cases normal bases
along $\gamma$ always exist (respectively at a point or along a path), which
was already established in~\cite{f-Frames-n+point,f-Bases-n+point} (and
independently in~\cite{Hartley}) and in ~\cite{f-Frames-path} respectively.

	In  the  other  limiting case, $p=n:=\dim (M)$, it is easily seen
that  the quantities~(\ref{14})  are simply the matrices formed from the
components of the corresponding  curvature
tensor~\cite{f-Frames-n+point,K&N-I,Lovelock-Rund} and that the  set
$\gamma (J^{p})$ consists of one or more neighborhoods in $M$. Consequently,
now theorem~\ref{Th4} states that the normal frames investigated here
exist iff the corresponding derivation is flat, i.e. if its curvature tensor
is zero, a result already found in~\cite{f-Frames-n+point}.

	In the general case, when $2\le p<n$ (for $n\ge 3$), normal bases,
even anholonomic,  do not exist if (and only if)  the conditions~(\ref{22})
are not satisfied. Besides, in this case the quantities~(\ref{14}) cannot be
considered as a `curvature' of $\gamma (J^{p})$. They  are something like
`commutators' of covariant derivatives of  a type $\nabla _{F}$,  where $F$
is a tangent to $\gamma (J^{p})$ vector field
(i.e.  $\left.F\right|_x\in \left.T\right|_x(\gamma (J^{p}))$
if $\gamma (J^{p})$ is a submanifold of $M)$,
and which act on tangent to $M$ vector fields.

	Let us also note that the normal frames on a set $U$  are  generally
anholonomic. They may be holonomic  only in the torsion free case  when the
derivation's torsion vanishes on $U$.

	The results of this work, as well as the ones
of~\cite{f-Frames-n+point,f-Frames-path}, are important in connection with
the use of normal frames in gravitational
theories~\cite{MTW,Gronwald/Hehl-96}.
In particular now we know that there exist normal frames (at a point or along
paths) in Riemann-Cartan spacetimes, a problem that was open until
recently~\cite{Gronwald/Hehl-96}.

	The above results outline the general bounds of  validity  and
express the exact mathematical form of the  equivalence  principle.  This
principle requires~\cite{MTW} that the  gravitational  field  strength,
theoretically identified  with  the components of a linear connection, can
locally  be transformed  to zero by a suitable choice of the local reference
frame (basis),  i.e. by it  there have to exist local bases in which the
corresponding connection's components vanish.

	The above discussion, as well as the results
from~\cite{f-Frames-n+point,f-Frames-path}, show the identical validity of
the equivalence principle in zero and one dimensional cases, i.e.  for $p=0$
and $p=1$. Besides, these are the \emph{only cases} when  it is fulfilled for
\emph{arbitrary} gravitational fields.  In fact, for $p\ge 2$ (in the case
$n\ge 2$), as we saw in Sect.~\ref{IV}, normal bases do not exist unless the
conditions~(\ref{22}) are satisfied.  In particular,  for $p=n\ge 2$ it is
valid only for flat linear connections (cf.~\cite{f-Frames-n+point}).

	Mathematically the equivalence principle is expressed through
corollary~\ref{Cor10} or, in some more general situations, through
theorem~\ref{Th4}.  Thus  we  see that in gravitational theories based on
linear connections this principle is identically satisfied at any fixed point
or along  any fixed path, but on submanifolds of dimension greater or equal
two it  is generally not valid.  Therefore in this class of gravitational
theories  the equivalence  principle is a theorem derived from their
mathematical background. It may play a role as a  principle if one tries to
construct a gravitational  theory  based  on  more general derivations, but
then, generally, it will reduce such  a  theory to one based on linear
connections.

	A comprehensive analysis of the equivalence principle on the base of
the present work and~\cite{f-Frames-n+point,f-Frames-path} can be found
in~\cite{f-PE-P?}.


\bibliography{bozhopub,bozhoref}
\bibliographystyle{unsrt}

\end{document}